\documentstyle[12pt,aaspp4,flushrt]{article}

\newcommand{\be}{\begin{equation}}
\newcommand{\ee}{\end{equation}}
\newcommand{\ltsima}{$\; \buildrel < \over \sim \;$}
\newcommand{\simlt}{\lower.5ex\hbox{\ltsima}}
\newcommand{\hii}{{H\sc{ii}~}}

\begin{document}

\title{SELF-SIMILAR CHAMPAGNE FLOWS IN \hii REGIONS}

\author{Frank H. Shu}
\affil{National Tsing Hua University, \\
		101 Section 2 Kuang Fu Road \\
		Hsinchu, Taiwan 30013, ROC\\
              shu@astron.berkeley.edu}

\author{Susana Lizano}
\affil{Instituto de Astronom\'{\i}a, UNAM \\
	      Campus Morelia \\
              Apdo 3-72 (Xangari)\\
              58089 Morelia, Michoac\'an, M\'exico\\
              s.lizano@astrosmo.unam.mx}

\author{Daniele Galli}
\affil{Osservatorio Astrofisico di Arcetri \\
              Largo Enrico Fermi 5 \\
              I-50125 Firenze, Italy \\
              galli@arcetri.astro.it}

\author{Jorge Cant\'o}
\affil{Instituto de Astronom\'{\i}a, UNAM \\
              Apdo 70-264 \\
              4510 M\'exico, D.F., Mexico}

\author{Gregory Laughlin}
\affil{Lick Observatory \\
	University of California at Santa Cruz\\
	Santa Cruz, CA 95064\\
        laugh@ucolick.org}

\begin{abstract}

We consider the idealized expansion of an initially self-gravitating,
static, singular, isothermal cloud core.  For $t\ge 0$, the gas is
ionized and heated to a higher uniform temperature by the formation of
a luminous, but massless, star in its center.  The approximation that
the mass and gravity of the central star is negligible for the
subsequent motion of the \hii region holds for distances $r$ much
greater than $\sim 100$ AU and for the massive cloud cores that give
rise to high-mass stars.  If the initial ionization and heating is
approximated to occur instantaneously at $t=0$, then the subsequent
flow (for $r \gg 100$ AU) caused by the resulting imbalance between
self-gravity and thermal pressure is self-similar.  Because of the
steep density profile ($\rho \propto r^{-2}$), pressure gradients
produce a shock front that travels into the cloud, accelerating the gas
to supersonic velocities in what has been called the ``champagne
phase.'' The expansion of the inner region at $t > 0$ is connected to
the outer envelope of the now ionized cloud core through 
this shock whose strength depends on the temperature of the \hii gas.
In particular, we find a modified Larson-Penston (L-P) type of solution
as part of the linear sequence of self-similar champagne outflows.  The
modification involves the proper insertion of a shock and produces the
right behavior at infinity ($v \rightarrow 0$) for an outflow of
finite duration, reconciling the long-standing conflict on the correct
(inflow or outflow) interpretation for the original L-P solution.

For realistic heating due to a massive young central star which ionizes
and heats the gas to $\sim$ 10$^4$ K, we show that even the
self-gravity of the ionized gas of the massive molecular cloud core can
be neglected.  We then study the self-similar solutions of the
expansion of \hii regions embedded in molecular clouds characterized by
more general power-law density distributions: $\rho \propto r^{-n}$
with $ 3/2 < n < 3$.  In these cases, the shock velocity is an
increasing function of the exponent $n$, and diverges as $n \rightarrow
3$.  We show that this happens because the model includes an origin,
where the pressure driving the shock diverges because the enclosed
heated mass is infinite. Our results imply that the continued
photoevaporation of massive reservoirs of neutral gas (e.g.,
surrounding disks and/or globules) 
nearby to the embedded ionizing source is required
in order to maintain over a significant timescale the emission measure
observed in champagne flows.

\end{abstract}


\section{Introduction }
\label{intro}

For a spherically symmetric molecular cloud core, initially at rest, 
the size $r_S$ of the region that can be ionized, is given by the standard
formula (Str\"omgren 1939):
\be
\int_{r_0}^{r_S} n_e n_p \alpha_2 4 \pi r^2 dr = \dot N_\ast.
\label{rs}
\ee
Eq.~(\ref{rs}) assumes ionization equilibrium and the ``on the spot''
approximation.  In the above, $n_e$ is the electron density; $n_p$ is
the ion density; $\alpha_2$ is the recombination coefficient to the
second energy level of hydrogen; $\dot N_\ast$ is the rate of ionizing
photons from the star, assumed to be a constant; and  $r_0$ is the
radius below which all of the gas in the original cloud core may be
considered to have fallen into the center (perhaps via a disk) to make
a star of mass $M_\ast$.\footnote{In the case when $\dot N_\ast \propto
t^3$, Newman \& Axford (1968) found self-similar solutions for the
expansion of an ionization bounded \hii region in a uniform H{\sc I}
cloud.}.  If the virial velocity (thermal, turbulent, or
magnetohydrodynamic) supporting the original (neutral) cloud core
before star formation is denoted by $a_1$, order of magnitude arguments
yields $r_0 \sim r_1$, the Bondi-Parker radius of this
neutral gas,
\be
r_1 \equiv {GM_\ast \over 2a_1^2}.
\label{Bondione}
\ee
The square of the sound speed in the \hii gas $a_2^2$ is generally
appreciably larger than $a_1^2$; thus, the equivalent Bondi-Parker
radius of the ionized gas,
\be
r_2 \equiv {GM_\ast \over 2 a_2^2},
\label{Bonditwo}
\ee
will be considerably smaller than $r_1$.

For typical numbers, $M_\ast \simeq 25 \; M_\odot$, $a_1 \simeq 1$ km
s$^{-1}$, $a_2 \simeq 10$ km s$^{-1}$, we have $r_1 \simeq
10^4$ AU $\gg r_2 \simeq 10^2$ AU, with both $r_1$ and $r_2$ much
bigger than the physical radius of the star.  Much interior to $r_2$,
the ionized gas will empty into the star (or more likely, into a disk
if it has even a slight amount of angular momentum); whereas much
exterior to $r_2$, the gravitationally unbound \hii gas will expand
outward, if it has not already reached pressure equilibrium with the
surrounding cloud.  Since $r_0 \gg r_2$, we may henceforth ignore the
gravitational field of the star on the flow of the \hii region beyond
$r_0$, although for purposes of making contact with earlier theoretical
work, we shall begin by not ignoring the self-gravity of this gas.
Since the material inside $r_0$ of the initial density profile should
have fallen into the star, the observed presence of appreciable amounts
of ionized gas at intermediate radii, $\sim 10^3$ AU in typical
ultracompact \hii regions, is awkward to explain.  We defer until \S 6
the discussion of the special kinds of models that are probably required
to explain ultracompact \hii regions.

Assume now that the molecular cloud core initially had
a power-law distribution of gas density that extends essentially
to infinity:
\be
\rho(r) = K r^{-n}.
\label{powerlaw}
\ee
If $n < 3/2$, the ultraviolet radiation is trapped within a finite
radius $r_S$, and the \hii region is said to be ``ionization bounded''
(see Osterbrock 1989).  If $n > 3/2$, the \hii region can be either
ionization bounded or ``density bounded.''  In the latter case, a
finite output of ultraviolet radiation can ionize an infinite volume of
gas beyond $r_0$.  The dividing line between being ionization bounded
and density bounded arises when the density constant $K$ equals a
critical value $K_{\rm cr}$:
\be
K_{\rm cr} =  2 \mu_i m_H \left[ { {(2n-3) r_0^{2n-3} \, \dot N_\ast}
\over{ 4 \pi \alpha_2 } }\right ]^{1/2},
\label{Kcrit}
\ee
where $\mu_i$ is the mean weight per particle of the ionized gas,
$m_H$ is the hydrogen mass, and
$n_p=n_e=\rho/2 \mu_i m_H $.

We wish to compare the value of $K_{\rm cr}$ with the value $K_\ast$
implied by the assumption that the power law (\ref{powerlaw}) initially
extended inward from $r_0$ as well as outward, but that the
part inward of $r_0$ has fallen into the center (perhaps via a disk)
to make a star of mass $M_\ast$:
\be
K_\ast = {(3-n)M_\ast\over 4\pi r_0^{3-n}}.
\label{Kstar}
\ee
Taking the ratio of eq. (\ref{Kstar}) to eq. (\ref{Kcrit}),
we get
\be
{K_\ast\over K_{\rm cr}} = \left[{(3-n)M_\ast\over 2\mu_i m_H}\right]
\left[{\alpha_2\over (2n-3)4\pi r_0^3\dot N_\ast}\right]^{1/2}.
\label{Kratio}
\ee
For $M_\ast \simeq 25~M_\odot$, $r_0 \simeq 10^4$~AU, $\dot N_\ast \simeq 
10^{49}$ s$^{-1}$, $\alpha_2 \simeq 2.6 \times 10^{-13}$~cm$^3$ s$^{-1}$,
$K_\ast/K_{\rm cr} \simeq 23\, (3-n)/(2n-3)^{1/2}$.

It is remarkable that factors of such disparate orders of magnitude as
the dimensionless quantities in the two square brackets of eq.
(\ref{Kratio}) combine to give a ratio within two orders of unity.
Nevertheless, since $K_\ast$ represents a rough estimate of $K$ and
$K_\ast > K_{\rm cr}$, this calculation formally indicates that the
\hii regions of 25 $M_\odot$ (and lower mass) stars are likely to be
ionization bounded, at least initially before any expansion occurs.
However, if we assume that $\dot N_\ast$ scales roughly as $M_*^3$ 
(as indicated by the results of
Vacca et al.~1996), the expression on the right-hand side scales as
$M_\ast^{-2}$, indicating that the \hii regions of the most massive O
stars may be density bounded from the start, especially if such stars
are born in regions with density gradients close to $n=3$.  They will
then develop champagne flows as follows.

When $K \sim K_\ast < K_{\rm cr}$, the ionization front (IF) created by
the idealized instantaneous appearance of a star at $t=0$ rapidly moves
to infinity and establishes an isothermal structure with $T \simeq
10^4$ K.  After the passage of the IF, the cloud remains out of
mechanical balance and the pressure gradients will produce an expansion
of the whole cloud. Due to the density gradient the inner regions
expand faster than the outer regions and a shock travels through the
cloud, accelerating the gas to supersonic velocities. This is known as
the ``champagne phase'' (e.g.  Bodenheimer, Tenorio-Tagle \& Yorke
1979).  Franco, Tenorio-Tagle \& Bodenheimer (1990; hereafter FTB)
studied the evolution of \hii regions embedded in molecular clouds with
steep density gradients.  High spatial resolution infrared and radio
recombination line observations toward several sources have found
ionized gas accelerating away from the central source in the manner
expected of champagne flow models (e.g. Garay et al. 1994; Keto et al.
1995; Lumsden \& Hoare 1996; Lebr\'on et al. 2001).  Note that in
several of the observed compact \hii regions 
(e.g., 29.96-0.02, G32.80+0.19B, G61.48+0.09B1) the inferred rate of
ionizing photons imply excitation by central stars with masses $M_* > 30 M_\odot$.

Density profiles in massive molecular cores have also been extensively
studied observationally (e.g.  Garay \& Rodr\'\i guez 1990; Caselli \&
Myers~1995; Van der Tak et al. 2000; Hatchell et al. 2000; for a review
see Garay \& Lizano 1999). Even though the environment is possibly
clumpy on scales of tenths of pc, density profiles are well
approximated by power laws with $1 \simlt n \simlt 2$.  Theoretical
models of the formation of massive stars within dense and massive cores
have assumed power law exponents in this range (Osorio et al. 1999;
McKee \& Tan 2002).  Recently, Franco et al. (2000) have argued that
radio continuum spectra of ultracompact \hii regions indicate initial
density gradients with $2 \simlt n \simlt 3$.  Clearly, more
observations with high spatial resolution are necessary to reliably
establish the density profiles of the sites of massive star formation.

The purpose of this paper is to study by similarity techniques the
``champagne phase'' of expansion of \hii regions with power law density
distributions.  In \S 2, we formulate the outflow problem in the case
of the singular isothermal sphere (SIS) that has $\rho \propto r^{-2}$,
including the effect of self-gravity. Tsai \& Hsu (1995) found the
outflow analogue of the inside-out collapse solution (Shu 1977), but in
which the SIS is sent into expansion by an outward propagating shock.
In \S 3 we show that the Tsai \& Hsu (1995) solution is actually the
limit of a family of outflow solutions when $(a_1/a_2)^2 \rightarrow 1$
from below.  Furthermore, the outflow solution with the particular
ratio of $(a_1/a_2)^2 =0.75$ corresponds to a piece of the
time-reversed Larson-Penston (L-P) collapse solution (Larson 1969;
Penston 1969), but with a shock inserted to obtain the correct
asymptotic behavior for large distances (or early times). For realistic
heating after the passage of an IF, i.e., for realistic values of
$(a_1/a_2)^2 \ll 1$, we show that the self-gravity of the \hii gas can
be neglected.  In \S 4 we extend our study to the evolution of
champagne flows in the case of density distributions with power law
exponents in the range $3/2 < n < 3$, neglecting self-gravity.  In \S 5
we find that the self-similar models have a shock propagating at
constant velocity into the ionized gas, in good detailed agreement with
the models of FTB. In particular, the shock velocity diverges as $n
\rightarrow 3$.   We show that this happens because the formal
treatment extends the inner radius of the calculation to the origin.
In such a treatment, the mass of driving \hii gas diverges when $n \ge
3$.  We perform more realistic calculations in such cases that cut
holes in the gas distribution for $r < r_0$.  In \S 6 we summarize our
conclusions, and we discuss the implications of our results for the
problem of ultracompact \hii regions. Finally, in the appendices we
show that, for the scales relevant to molecular cloud cores, the
isothermal assumptions for the gas and the shock are valid.

\section{Governing Equations}
\label{ss}

Consider a star forming cloud core with
the density profile of the SIS
\be
\rho(r) ={ a_1^2 \over 2 \pi G r^2},
\ee
where $a_1$ is the sound speed of the cloud at $t < 0$.  Imagine that
at $t=0$, the central star turns on and heats the entire cloud core to
a uniform temperature appreciably higher than it had originally
(perhaps by the rapid propagation of an ionization front).  Let $a
\equiv a_2 > a_1$ be the sound speed corresponding to this new
temperature. In order to keep the initial gas density distribution
unchanged from $t= 0^-$ to $t=0^+$, it is convenient to write
\be
\rho(r,0^+) = {\epsilon a^2 \over 2 \pi G r^2},
\label{tplus}
\ee
where $\epsilon\equiv (a_1/a)^2\leq 1$.  The original temperature was
appropriate for a self-gravitating SIS, but the new higher temperature
makes the same density distribution over-pressured. 
The gas expands and a shock front propagates through the cloud, 
setting it into a champagne outflow.

The expansion of self-gravitating champagne flows is governed by the
continuity equation,
\be
{\partial \rho  \over \partial t} 
+ {1 \over r^2} {\partial \left( r^2 \rho u \right) \over \partial r} =0, 
\label{cont}
\ee
where $\rho$ is the gas density and $u$ is the gas
velocity, and by the momentum equation,
\be
{\partial u \over \partial t} 
+ u {\partial u  \over \partial r} = 
-{a^2 \over \rho} {\partial \rho \over \partial r}
 - {G M \over r^2},
\label{moment}
\ee
where $G$ is the gravitational constant and $M$ is the mass of the gas
inside the radius $r$.  An energy equation is not required since we we
have assumed an isothermal equation of state $P=a^2 \rho$, with shock
jumps also occurring isothermally (see Appendix~B).

We further simplify the treatment by approximating the central star as
a source of light and heat (to keep the gas ionized and warm), but not
of mass, and we extend the range of the flow calculation right into the
origin.  We justify this cavalier treatment of the complex situation
interior to $r=r_0$ by noting that the first thing all champagne
solutions try to establish is a central region of quasi-uniform
pressure (see \S\S 3 and 4).  All the available ionized mass interior
to $r=r_0$ is homogenized by the high pressure in less than a sound
crossing time (because supersonic motions are involved), $r_0/a$.
Thus, all detail about the structure of the region interior to $r=r_0$
is lost in a time $\le 5\times 10^3$ yr, which is quite short compared
to the duration of interest for champagne flows.   Apart from initial
transients, then, the evolution of quasi-uniform \hii regions of high
over-pressure expanding into exterior regions with power-law density
profiles should follow the description given by the self-similar
treatment of this paper.  An exception to this statement arises if the
region interior to $r=r_0$ is a continuous source of additional ionized
gas (derived, say, from the {\it continuous} rather than instantaneous
photoevaporation of a circumstellar disk and/or remnant neutral globules).  
We ignore this exception
for the present, but we return to the possibility in \S6 when we
comment on the problem of ultracompact \hii regions.

\subsection{Similarity assumption}

Following Shu (1977) we introduce the similarity
variable
\be
 x = {r \over a t},
\label{n1}
\ee
and we define the reduced density
\be
\rho(r,t) = {\alpha(x) \over 4 \pi G t^2},
\label{n2}
\ee
the reduced mass
\be
M(r,t) = {a^3 t \over G} m(x),
\label{n3}
\ee
and the reduced velocity 
\be
u(r,t) =a v(x).
\label{n4}
\ee
%Note that the SIS has $\alpha = 2/x^2$, $m=2 x$ and $v=0$.

%Changing variables $(r,t)$ to $(x,t)$ implies that
%$ \left. {\partial \over \partial t} \right \vert_r 
% = \left. {\partial \over \partial t} \right \vert_x - 
%{x \over t} \, \left. {\partial \over \partial x} \right \vert_{t} $, and
%$ \left. {\partial \over \partial r} \right \vert_t = {1 \over {at}} 
% \left. {\partial \over \partial x} \right \vert_{t} $.
Substituting these expressions in eq.~(\ref{cont}) and (\ref{moment}), 
one obtains two coupled first order differential equations (ODEs)
for the reduced density $\alpha$ and velocity $v$
\be
[ (v - x)^2 - 1]{1 \over \alpha} {d \alpha \over d x}  =
 \left[\alpha - {2 \over x}(x - v) \right](x-v),
\label{alpha}
\ee
\be
[ (v - x)^2 - 1]{d v \over d x} =
 \left[(x- v)\alpha - {2 \over x} \right](x-v), 
\label{vel}
\ee
while the reduced mass is given by
\be
m = x^2 \alpha (x -v).
\label{mass}
\ee

We identify $t=0$ as the initial instant and require $t, \ x, \ m, \ \alpha$ 
and $v$ all to be positive.  Then, the critical lines
where the LHSs and RHSs of eq.~(\ref{alpha}) and (\ref{vel}) vanish 
lie on the locus
\be
v_c = x_c -1, \quad\quad\quad \alpha_c = {2 \over x_c}.
\label{critical}
\ee
A special case with an analytical solution that passes smoothly
through the critical line at $x_c=3$ was cited by Whitworth \& Summers (1985;
eq. [3.5]):
\be
\alpha = {2\over 3}, \qquad v ={2\over 3} x.
\ee
In dimensional form, this is the model for the Einstein-De Sitter
universe (in the non-relativistic limit where $a^2 \ll c^2$ so that we
can ignore the contribution of the gas pressure in determining
spacetime curvature), with a Hubble ``constant'' and 
cosmic mass density given respectively by
$H = 2/(3t)$ and $\rho = 3H^2/(8\pi G) = 1/(6\pi G t^2)$.

In the \hii region problem, the reduced central density
$\alpha(0) \equiv \alpha_0$ will not usually be
tuned to the specific value $\alpha_0 = 2/3$ required
to make a smooth crossing of the critical line.  Even more,
the pressure (and density) homogenization of the central
regions cannot extend instantaneously to all space because
of the existing density gradient at $t=0$.
Nevertheless, if we ignore the central star (or more accurately, if we
put it back to fill the hole inside $r_0$), the behavior
of the Einstein-De Sitter solution is generic for the inner
regions of the \hii region in that the dimensional central density
will decline with time as $t^{-2}$  because
of the expansion of the flow toward the exterior regions
of lower pressure and density.
Thus, for arbitrary values of $\alpha_0$, we must solve
eq.~(\ref{alpha}) and (\ref{vel}) subject to the inner BCs:
\be
\alpha = \alpha_0 \quad\mbox{and}\quad v = 0, \quad\mbox{at}\quad x = 0.
\label{BCd}
\ee
A series expansion near the origin (a regular singular point
of eqns.~[\ref{alpha}] and [\ref{vel}]) now yields 
\be
\alpha=\alpha_0+{\alpha_0\over 6}\left({2\over 3}-\alpha_0\right)x^2+\ldots,
\ee
\be
v={2\over 3}x+{1\over 45}\left({2\over 3}-\alpha_0\right)x^3+\ldots.
\label{seriesexp}
\ee

The heating of the cloud at $t=0$ introduces an imbalance between
self-gravity and pressure that will induce the propagation
of a shock and an outward subsonic flow of the
entire system for $t > 0$.
We assume isothermal jump conditions
\be
(v_u - x_s)(v_d - x_s) = 1, \quad {\alpha_d \over \alpha_u} = (v_u - x_s)^2,
\label{jump}
\ee
where the subscript $u$ ($d$) indicates the value of the reduced velocity
and density upstream (downstream) of the shock and $x_s$ is the position
(and velocity) of the shock in similarity coordinates. From eq.~(\ref{n1})
the position and velocity of the
shock in physical space are $r_s = x_s a t$ and
$ u_s \equiv d r_s/ d t = a x_s$.

The outer solution to which this shock attaches is given
by one of the ``plus solutions'' of Shu (1977), and it has the 
asymptotic behavior
\be
\alpha \rightarrow {2\epsilon \over x^2}, \quad 
v \rightarrow {2(1-\epsilon)\over x}, \quad {\rm as} \quad x \rightarrow \infty,
\label{BCup}
\ee
with $\epsilon \leq 1$.

\subsection{Numerical procedure}

We start the numerical integration of the downstream flow
at small $x$, using the second-order expansions (\ref{seriesexp})
for a given value of $\alpha_0$. At each integration point, we apply the
isothermal jump conditions and a provisional upstream solution is
obtained. A consistent solution is found when the upstream
solution, integrated to large $x$ ($\sim 10^3$), fulfills the
asymptotic BCs (\ref{BCup}). Thus, for a given $\alpha_0$ there is a
corresponding unique value of $\epsilon$ such that the full solution
satisfies the BCs (\ref{BCd}) and (\ref{BCup}) and the jump conditions,
eq.~(\ref{jump}).

\section{Outflow Solutions for the SIS}

By varying the value of $\alpha_0$, we obtain a family of outflow
solutions.  
For increasing values of $\alpha_0$, the shock velocity and
shock strength decrease, and the shock attaches to outer solutions
characterized by increasing values of $\epsilon\leq 1$.
Figure~\ref{fig1} shows the reduced velocity $v$ and density $\alpha$
versus $x$ for $\alpha_0 = 2/3$, 1, 1.67, 4, and 7.90,
corresponding to $\epsilon=0.545$, 0.640, 0.755, 0.916, and 1.

\subsection{The Tsai \& Hsu solution}

There is a maximum value $\alpha_0 = 7.90$ which corresponds to the
solution that attaches upstream to the static unperturbed SIS solution
with $\epsilon=1$ (thick solid lines in Figure~\ref{fig1}). In this
case the shock moves at $u_s = 1.34 a$, setting the unperturbed SIS
into motion. Since $\epsilon=1$, the temperature of the cloud has not
changed at $t=0$. Thus, the shock is driven by an infinitesimal
pressure pulse at the origin and acquires finite strength as it races
down a density gradient.  This self-similar solution, found first by
Tsai \& Hsu (1995), is the outflow counterpart of the ``expansion wave
solution'' (Shu 1977) for the collapse of the SIS, but now the SIS is
sent into expansion by an outwardly propagating shock.  Thus, the SIS
is unstable not only to self-similar ``inside-out collapse'' but also
to ``inside-out expansion'' (Tsai \& Hsu 1995).  (Of course,
physically, there is no reason to expect a perturbational heating of
the cloud core without having, first, inward collapse to form a
central star.) This special outflow solution is the limit of such
solutions as $\epsilon \rightarrow 1$.

\subsection{The Larson-Penston solution}

\begin{figure}
\plotone{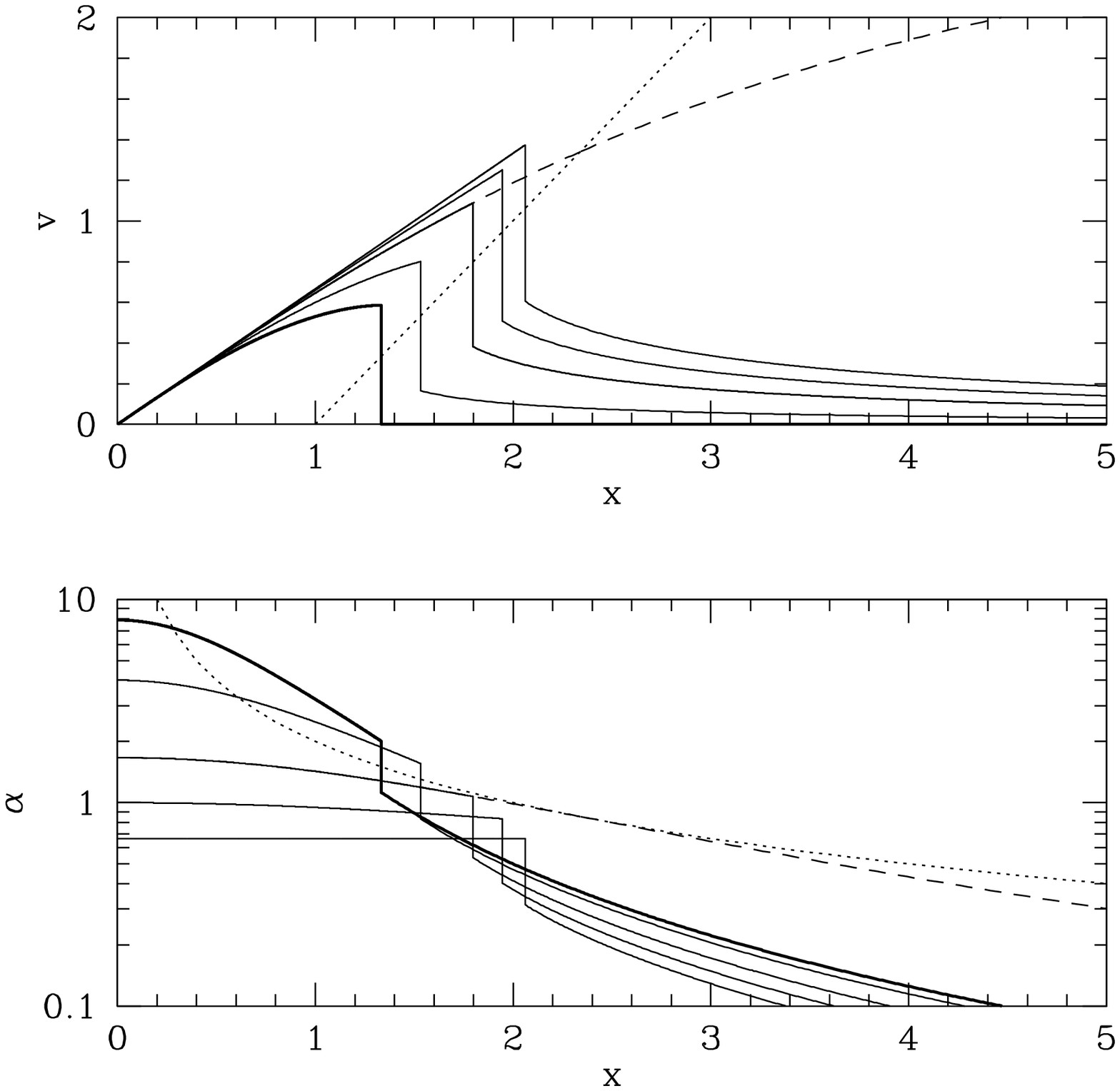}
\caption{
{\it Top panel:} reduced velocity $v$ for the outflow solutions of the
SIS, including self-gravity. The different curves from top to bottom
correspond to $\alpha_0 = 2/3$, 1 , 1.67, 4 and 7.90 ({\it thick
solid curve}\/).  The {\it dotted line} is the critical line $v = x -
1$, {\it Bottom panel:} reduced density $\alpha$ for the same values of
$\alpha_0$.  The {\it dotted curve} is the critical line $\alpha = 2/
x$.  In both panels, the {\it long-dashed curves} show the L-P solution
without a shock.  The parameters of the solutions are described in Table~1.}
\label{fig1}
\end{figure}

For the particular value $\alpha_0=1.67$ the flow downstream from the
shockfront is coincident with a time reversed piece of the L-P solution
originally proposed to describe infall.  Without introducing a shock,
the L-P solution crosses smoothly the critical lines at $x_c =2.34$
(dotted lines in Figure~\ref{fig1}), the reduced density being tangent
to the critical line, and achieves the asymptotic behavior
\be
\alpha \rightarrow {8.85 \over x^2} \quad\mbox{and}\quad v \rightarrow 3.28, 
\quad\mbox{for}\quad x \rightarrow \infty .
\ee
The L-P solution is represented by dashed lines in Figure~\ref{fig1}.
In contrast, our outflow solution has a shock at $x_s = 1.80$ 
and has the asymptotic behavior
\be
\alpha \rightarrow {1.50 \over x^2} \quad\mbox{and}\quad v \rightarrow 0,
\quad\mbox{for}\quad  x \rightarrow \infty ,
\ee
which is less dense than the SIS at large $x$, as appropriate for an
outflow solution. In fact, as discussed by Shu (1977) the 
time reversed L-P solution does not represent a proper wind solution,
precisely because it is overdense at large radii
with respect to the hydrostatic SIS.

\subsection{The Einstein-De Sitter solution}

For the special case $\alpha_0 = 2/3$, the solution interior
to the shockwave at $x_s=2.06$ is the Newtonian analogue
to an Einstein-De Sitter ($\Omega=1$) universe.  The
pressure (and density, for an isothermal gas) homogenization
of the interior is perfect in this case, and Hubble's law of
expansion is also exactly satisfied before the gas flow reaches
the radius $r=x_s at$.  Upstream from the shock front, the flow
corresponds to a breeze (see next subsection)
blowing slowly through an $\epsilon=0.545$
singular isothermal sphere.  Notice that this use of
the Einstein-De Sitter interior solution (connection to an
exterior solution via a shockwave) differs from that shown in 
Fig.~1 of Hennebelle~(2001) (connection via a weak discontinuity).

Unlike its cosmological
application, the Einstein-De Sitter (inner) solution represents 
nothing special in the
problem of the self-similar isothermal expansion of self-gravitating
\hii regions.  It merely demarcates the case (see Table 1) when
the post-shock density $\alpha_d$ switches from values
lower than the central level $\alpha_0$ (for $\alpha_0 > 2/3$)
to values higher than the central level (for
$\alpha_0 < 2/3$).  This switch occurs with increasing
heating of the \hii gas relative to its neutral
precursor (smaller values of $\epsilon \equiv [a_1/a_2]^2$).
With weak heating, running down a
density gradient established by the self-gravitating equilibrium
of the pre-ionization state and modified by a slowly blowing, upstream
breeze counts for more than compression
of this gas by a relatively weak shockwave.
With strong heating, shockwave compression
gains in the competition relative to farther propagation
into regions of lower density.

\subsection{General Behavior}

With the exception of the ``inside-out expansion'' solution of Tsai \&
Hsu (1995), the upstream flows have $v_u > 0$. This is due to the
mechanical imbalance following the heating of the cloud at $t=0$
causing the gas to expand subsonically before the shock arrives.  Such
subsonic behavior is characteristic of a ``breeze'' solution in the
nomenclature of the solar-wind literature, and we have referred to it as
such.  We note, however, that the arrival of the shockwave
pushes the breeze into supersonic expansion relative to the origin.
The strength of the shock and the value of $v_u$ depend on the new
temperature of the cloud at $t=0^+$.  Table 1 summarizes the results
for the self-similar models with different values of the central
density $\alpha_0$.  The columns are: 1) the value of $\alpha_0$; 2)
the parameter $\epsilon$ of the upstream solution; 3) the position of
the shock, $x_s$; 4) the post-shock velocity $v_d$; 5) the post-shock
density $\alpha_d$; 5) the pre-shock velocity $v_u$; and 6) the
pre-shock density $\alpha_u$.
The entries in this table are approximated to three significant figures.
The rounding-off error affects the fulfillment of the jump conditions
at the level of $\simlt 2\%$, a sufficient accuracy for practical 
applications.

\begin{deluxetable}{llllllll}
\tablecaption{SIS Outflow Solutions as Function of $\alpha_0$}
\tablehead{ $\alpha_0$ & $\epsilon$ & $x_s$ & $v_d$ & $\alpha_d$ & $v_u$ & $\alpha_u$ \\}
\startdata
  7.90   & 1     & 1.34 & 0.586 & 2     & 0     & 1.12  \\
  4      & 0.916 & 1.53 & 0.801 & 1.56  & 0.168 & 0.835 \\
  1.67   & 0.755 & 1.80 & 1.09  & 1.07  & 0.389 & 0.539 \\
  1      & 0.640 & 1.95 & 1.25  & 0.833 & 0.513 & 0.404 \\
  2/3    & 0.545 & 2.06 & 1.37  & 2/3   & 0.605 & 0.315 \\
  0.1    & 0.168 & 2.42 & 1.77  & 0.173 & 0.905 & $7.49 \times 10^{-2}$ \\
  0.01   & $2.09 \times 10^{-2}$ & 2.55 & 1.90 & $2.05\times 10^{-2}$ & 1.00 & $8.64 \times 10^{-3}$ \\
  $10^{-3}$ & $2.15 \times 10^{-3}$ & 2.56 & 1.91 & $2.10\times 10^{-3}$ & 1.02 & $8.78 \times 10^{-4}$ \\
  $10^{-4}$ & $2.15 \times 10^{-4}$ & 2.56 & 1.91 & $2.10\times 10^{-4}$ & 1.02 & $8.80 \times 10^{-5}$ \\
  $10^{-5}$ & $2.15 \times 10^{-5}$ & 2.56 & 1.91 & $2.10\times 10^{-5}$ & 1.02 & $8.80 \times 10^{-6}$ \\
\enddata
\label{T1}
\end{deluxetable}

\subsection{The limit $\epsilon \rightarrow 0$}

As discussed in \S~\ref{ss}, when the cloud core is heated
instantaneously at $t=0$, the value of $\epsilon$ must be chosen for
$t=0^+$ so that the physical density $\rho(r,0)$ remains unchanged.
For a cloud heated by the passage of an ionization front when a massive
star turns on in the center of the core, one expects the equivalent
thermal speed to increase from $a_1 \simeq 1$ km s$^{-1}$ to $a \simeq
10$ km s$^{-1}$.  Thus, the value of $\epsilon$ for the upstream flow
should be quite small, $\epsilon \simeq 10^{-2}$.   Since the
equilibrium configuration of the cloud has $\epsilon=1$ for $t<0$, the
condition $\epsilon \ll 1$ implies that self-gravity is not very
important in the subsequent evolution of the cloud for $t>0$.

We find that as $\epsilon \rightarrow 0$, the reduced velocity $v$
converges to an invariant solution with the fastest and strongest
shock.  Table 1 shows that most flow properties have converged pretty
much to their limiting values when $\epsilon \simeq 10^{-2}$.  The only
formal exception is the reduced density, which converges to a shape
invariant solution that scales as $\alpha_0 \rightarrow 0.47 \,
\epsilon$.  Thus, in the limit $\epsilon \rightarrow  0$, we find it
convenient to re-scale the density as \be R(x) = {x^2 \alpha(x) \over
2\epsilon}.  \ee The physical density for $t>0$ in term of $R$ is given
now by \be \rho(r,t) = {\epsilon a^2\over 2 \pi G r^2} R(x).
\label{Rphys} \ee Comparison with eq.~(\ref{tplus}) shows that $R=1$ at
$t=0^+$.

\begin{figure}
\plotone{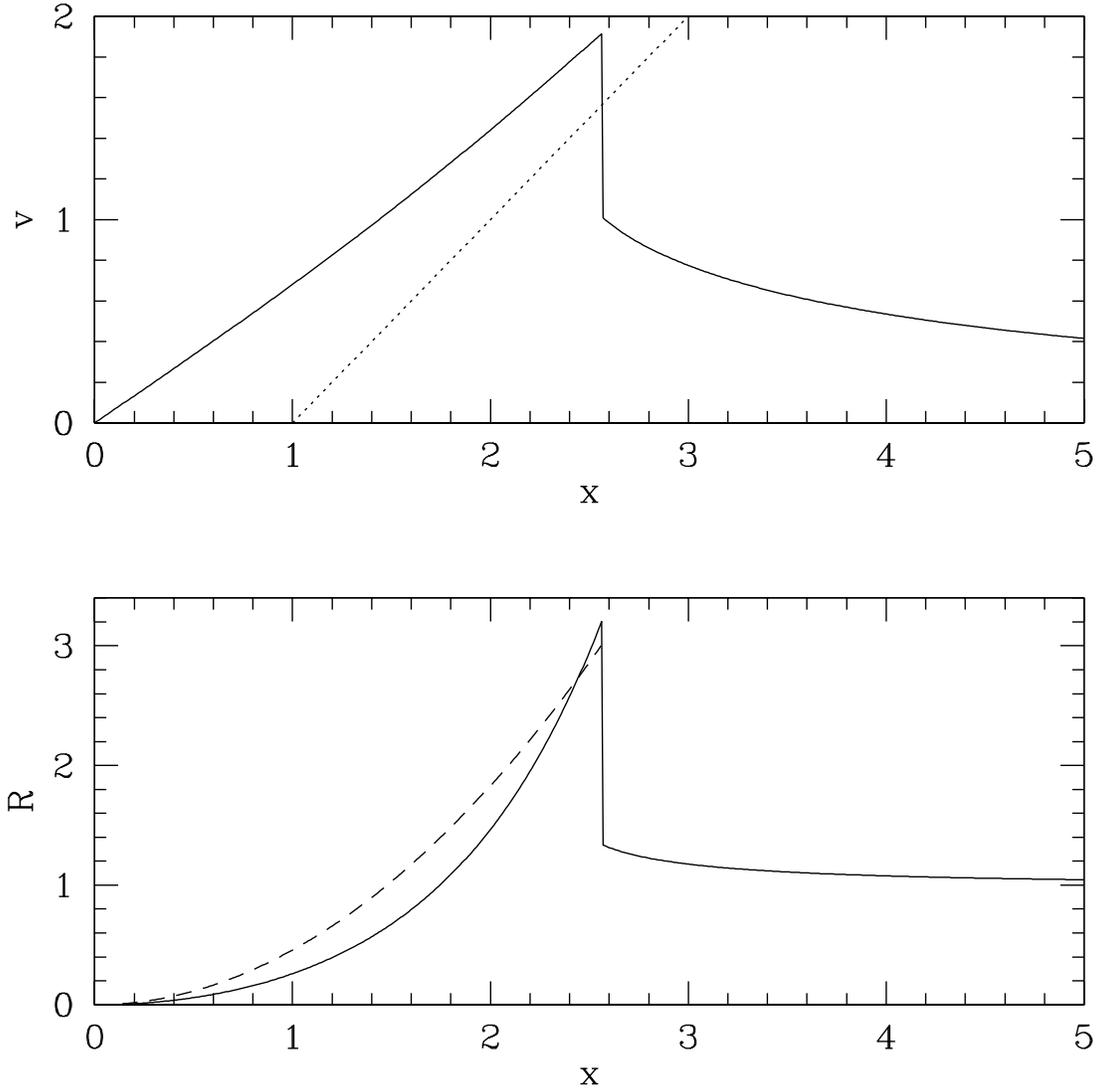}
\caption{
{\it Top panel:} reduced velocity $v$ for the exponent $n=2$, in
the limit $\epsilon = 0$. The {\it
dotted line} is the critical line $v = x - 1$.  The shock front is at
$x_s = 2.56$. The pre-shock velocity is $v_1 = 1.02$. {\it Bottom
panel:} reduced density $R$. The {\it dashed curve} corresponds to $R =
3 (x / x_s)^2$ and shows the deviation from uniform density. }
\label{fig2}
\end{figure}

The upper panel of Figure~\ref{fig2} shows the reduced velocity $v$ in
the limit $\epsilon =  0$.  The position of the shock front is
at $x_s = 2.56$ and the upstream reduced velocity, due to the cloud
general expansion, is $v_u = 1.02$.  The dotted line shows the locus of
the critical line  eq. (\ref{critical}).  The lower panel of Figure
\ref{fig2} shows the reduced density $R$. The dashed line shows the
function $R =3 (x / x_s)^2$, which corresponds (see eq. [\ref{Rphys}])
to a region of uniform but steadily decreasing density,
\be
\rho(t) \simeq \left( {3\over x_s^2}\right){\epsilon a^2\over 2\pi Gt^2},
\ee
given by spreading the original gaseous mass interior to $r_s=x_sat$
evenly over the enclosed spherical volume.  Except for a slight
increase of $R$ from its average interior value of 3 to the postshock
value $R_d = 3.20$ just downstream from the shockfront at $x=x_s$,
Figure~\ref{fig2} shows that the high gaseous pressure does a fairly
good job of ironing out pressure differences in the interior volume.

In summary, for a density distribution in mechanical equilibrium at $t
< 0$, which comes out of balance at $t=0^+$ because it has been heated
suddenly to a temperature much higher than the original value, one
expects that the effects of self-gravity can be neglected with respect
to thermal pressure. In this limit where $\epsilon \rightarrow 0$, we
verify explicitly that the self-similar solution converges to a unique
solution, equivalent to the gravitationless case, $G=0$.  Thus, in the
next section we will formulate the problem of self-similar champagne
flow solutions for power laws of the gas density distribution different
from $n=2$ also in the limit of negligible self-gravity .

\section{Generalized Formulation Ignoring Self-Gravity of the Ionized Gas}

In this section we discuss the champagne flows associated with
generalized power-law density-profiles with $\rho \propto r^{-n}$ in
the limit where the self-gravity of the cloud can be neglected.  For $t
< 0$, these density profiles result from equilibrium between the
self-gravity of the cloud core and other forces in addition to thermal
pressure (see e.g. discussion on polytropic clouds in Galli et al.
1999). We assume that, after heating at $t=0$, thermal pressure will
dominate over such forces in the evolution of the cloud.  The governing
equations are given by eq.~(\ref{cont}) and (\ref{moment}) setting
$G=0$. We choose the same self-similar variable and reduced velocity
defined by eq.~(\ref{n1}) and (\ref{n3}) but we generalize the density
profile of eq.~(\ref{powerlaw}) as
\be
\rho(r,t) = {K \over r^n} R(x).
\label{plaw}
\ee

The nondimensional equations are
\be
[ (v - x)^2 - 1] {d R \over d x} =
{R \over x}[v(v-x)(n-2) - n], 
\label{Rn}
\ee
\be
[ (v - x)^2 - 1]{d v \over d x} = {2 v \over x}  - n. 
\label{vn}
\ee

For $x \rightarrow \infty$ the boundary conditions are
$v \rightarrow 0$ and $R \rightarrow 1$. Then, the asymptotic expansions are
\be
v \rightarrow {n \over x}  \quad {\rm and} \quad
R \rightarrow 1 + { n (n-1) \over {2 x^2}} \quad {\rm for} \quad x\rightarrow \infty.
\ee

At the origin, 
\be
v\rightarrow {n \over 3}x \quad {\rm and } \quad R \propto x^n \quad {\rm for} \quad
x \rightarrow 0.
\ee
This boundary condition on $R$ implies that as $r \rightarrow 0$ the
density is uniform and is only a function of time $\rho \propto
t^{-n}$.  This is the expected behavior of the central zone of the \hii
region, where the sound crossing time is smaller than the expansion
time.

We apply the isothermal jump conditions (\ref{jump}) to connect the
upstream and downstream solutions.  Since the equation for the reduced
velocity is now decoupled from the equation for the reduced density, a
simple way to find the position of the shock front, $x_s$, is to
integrate numerically eq.~(\ref{vn}) outward from $x=0$, and integrate
inward from a large $x$. We check every point of the downstream
solution until the jump condition in the velocity (\ref{jump}) is
fulfilled.

Once the reduced velocity $v$ is known, the density equation 
(\ref{Rn}) can be integrated as
\be
R(x) = R_b \,
\exp{\, \int\limits_{x_b}^{x} {1 \over x^\prime}
{ {[v \left (v-x^\prime \right ) \left (n-2 \right ) -n ]}
\over {[\left ( v-x^\prime \right )^2 -1]} }
 dx^\prime }. 
\ee
For the outer solution, $x_b = \infty$
and $R_b = 1$. For the inner solution, $x_b = x_s$ and 
$R_b = R_d$,  where $R_d$ is the downstream reduced density evaluated
from the jump condition (\ref{jump}).

Before proceeding to discuss the self-similar champagne flows, let us
examine the case $n=3$.  In this case, eq.~(\ref{Rn}) and (\ref{vn})
have the analytic solution $R=Cx^3$ and $v=x$, where $C$ is an
arbitrary constant. The jump conditions eq.~(\ref{jump}) imply that
$x_s \rightarrow \infty$ as $v_d \rightarrow x$. 
Thus, in spatial coordinates the shock front,
$r_s = x_s a t$, and the shock velocity, $u_s  = a x_s$, go to infinity.

In dimensional variables, the analytic solution for the case $n = 3$
simply corresponds to a Hubble flow in a $\Omega = 0$ universe: $u =
r/t$, $\rho \propto t^{-3}$.  The propagation of the shockwave to
infinity leaves behind it an expanding \hii region of uniform but ever
decreasing density that looks like a {\it Little Bang}.  We will
discuss below the physical meaning (or lack of it) of this remarkable
result.

\section{Outflow Solutions for Different Gradients}

\begin{figure}
\plotone{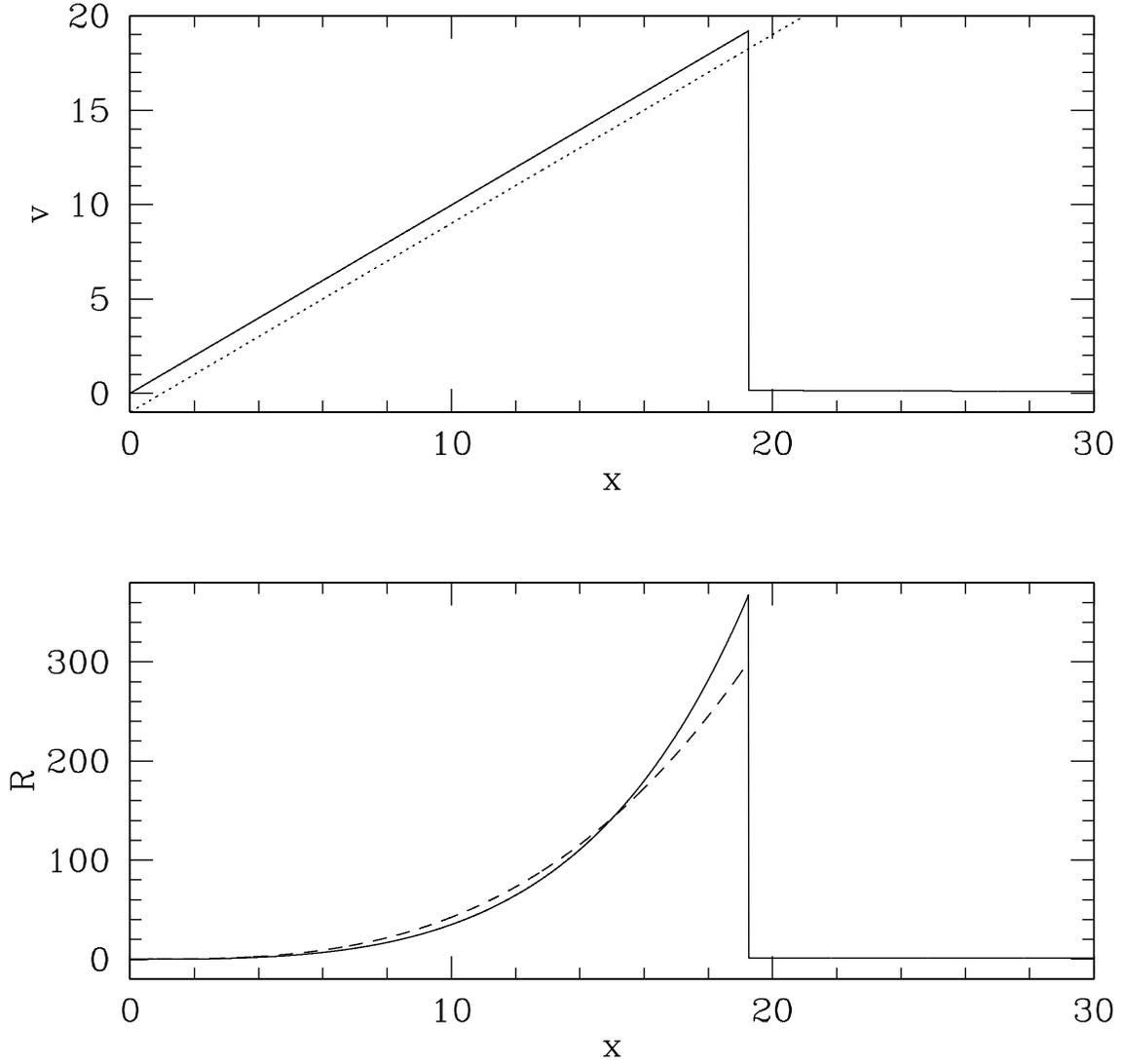}
\caption{
{\it Top panel:} reduced velocity $v$ for the exponent $n=2.99$. The
{\it dotted line} is the critical line $v = x -1$.  The shock front is
at $x_s = 19.25$. The pre-shock velocity is only $v_1 = 0.16$. {\it
Bottom panel:} reduced density $R$.  The {\it dashed line} corresponds
to $R = 300 (x / x_s)^{2.99}$ and shows the deviation from uniform density.}
\label{fig3}
\end{figure}

The nongravitating case for $n=2$ was shown in Figure~\ref{fig2}.
Figure~\ref{fig3} show the analogous reduced velocity $v$ and density
$R$ for $n=2.99$. The dotted line in the upper panel corresponds to the
critical line. In this case the shock front is at $x_s = 19.25$.   This
number is also the velocity of the shockwave relative to the
isothermal speed of sound of the \hii region.  In contrast, the
pre-shock velocity of the gas (relative to the origin) is only $v_u =
0.16 $, because at a given spatial position  there is less time for the
cloud to expand before the shock arrives. The reduced density $R$ in
the lower panel has a correspondingly large post-shock density
increase.  The dashed line shows the function
\be
R=\left( {3\over 3-n}\right)\left({x\over x_s}\right)^n,
\ee
given by spreading the original gaseous mass interior to $r_s=x_sat$
evenly over the enclosed spherical volume.  Figure~\ref{fig3} ~ shows
that the shock dynamics raises the immediate downstream value from the
average expectation $3/(3-n)=300$ at $x=x_s$ to the actual postshock
value $R_d = 368$.

Table~2 summarizes the results for the self-similar models with
different exponents $n$ in the density profile.  The columns are: 1)
the exponent $n$; 2) the position of the shock, $x_s$; 3) the
post-shock velocity $v_d$; 4) the post-shock density $R_d$; 5) the
pre-shock velocity $v_u$; and 6) the pre-shock density $R_u$.  One can
see that $x_s$ increases as $n \rightarrow 3$.

\begin{deluxetable}{lrrrrrr}
\tablecaption{Flow Parameters as a Function of $n$}
\tablehead{ $n$ & $x_s$ & $v_d$ & $R_d$ & $v_u$ & $R_u$ \\}
\startdata
    1.5  &     2.15 &   1.33  &  1.85  &   0.93  &  1.25  \\ 
    1.6  &     2.23 &   1.44  &  2.04  &   0.96  &  1.28  \\
    1.7  &     2.30 &   1.55  &  2.26  &   0.98  &  1.30  \\ 
    1.8  &     2.38 &   1.66  &  2.52  &   1.00  &  1.32  \\ 
    1.9  &     2.47 &   1.78  &  2.83  &   1.01  &  1.33   \\
    2.0  &     2.56 &   1.91  &  3.20  &   1.02  &  1.34   \\
    2.1  &     2.66 &   2.05  &  3.65  &   1.01  &  1.35   \\ 
    2.2  &     2.77 &   2.21  &  4.21  &   1.00  &  1.34   \\
    2.3  &     2.91 &   2.39  &  4.93  &   0.98  &  1.33   \\ 
    2.4  &     3.07 &   2.59  &  5.87  &   0.95  &  1.31   \\
    2.5  &     3.27 &   2.85  &  7.17  &   0.91  &  1.29   \\ 
    2.6  &     3.55 &   3.17  &  9.10  &   0.85  &  1.25   \\
    2.7  &     3.96 &   3.64  & 12.26  &   0.76  &  1.20   \\ 
    2.8  &     4.67 &   4.42  & 18.48  &   0.65  &  1.14   \\
    2.9  &     6.33 &   6.16  & 36.94  &   0.47  &  1.08   \\ 
    2.95 &     8.76 &   8.65  & 73.73  &   0.34  &  1.04   \\
    2.99 &    19.25 &  19.20  & 367.57 &   0.16 &   1.01   \\ 
\enddata
\label{T2}
\end{deluxetable}

\begin{figure}
\plotone{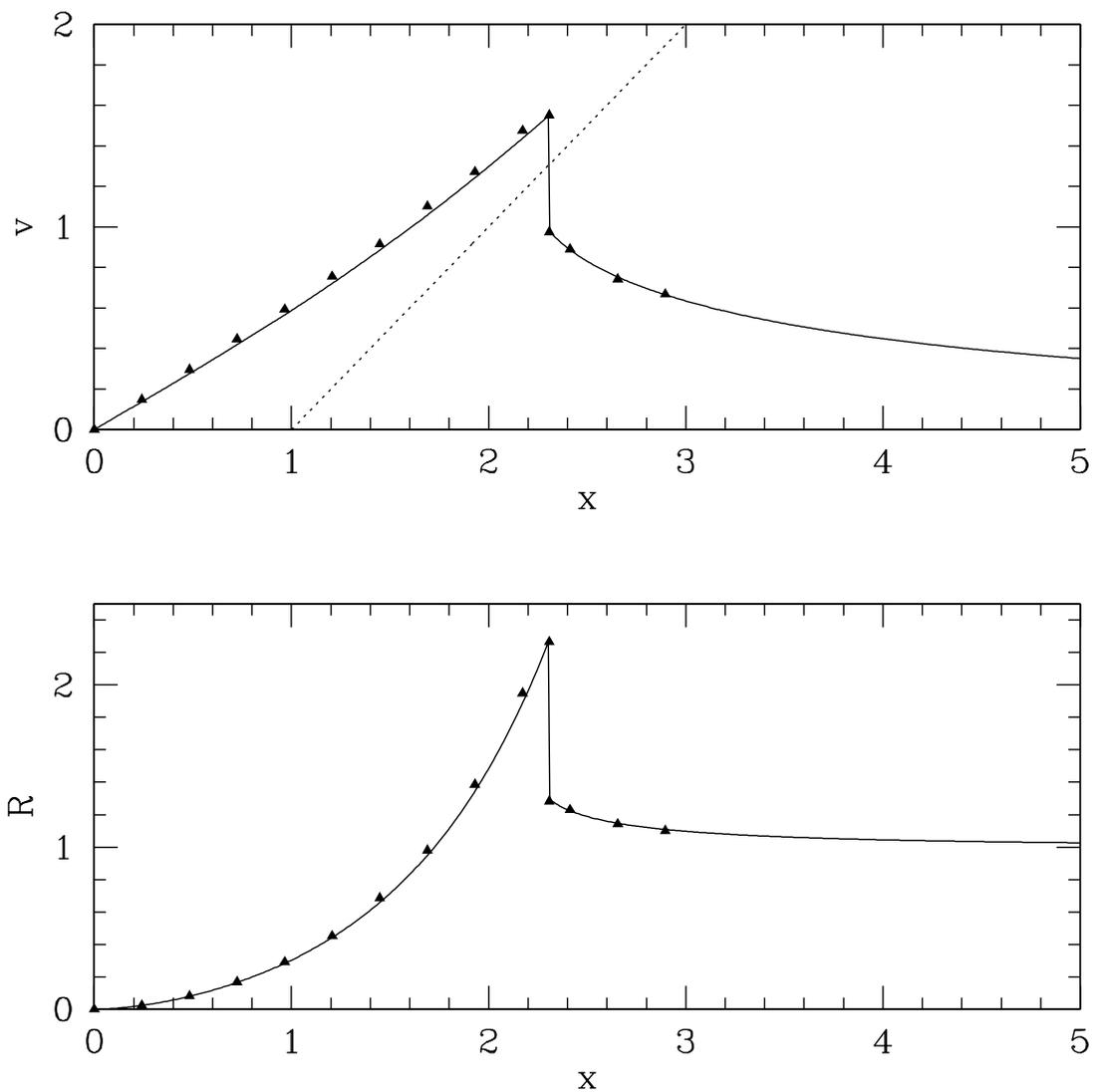}
\caption{
{\it Top panel:}  reduced velocity $v$ for the exponent $n=1.7$. The
{\it dotted line} is the critical line $v = x - 1$.  The shock front is
at $x_s =   2.30$. The pre-shock velocity is $v_u = 0.98$. {\it Bottom
panel:} reduced density $R$.  In both panels, the {\it triangles} show
the results of the numerical models of FTB for $n=1.7$ at $t=1.72
\times 10^5$ yr, expressed in our self-similar variables.}
\label{fig4}
\end{figure}

Figure~\ref{fig4} shows the comparison of our self-similar solutions
with the numerical results obtained by FTB in the case $n=1.7$. We
converted the values of the velocity and density calculated by FTB at
$t=1.72 \times 10^5$ yr (see their Figure~3c) into our self-similar
variables (eq.~\ref{n1}, \ref{n4} and \ref{plaw}) using their values of
the sound speed and their normalization for the density profile.
Figure~\ref{fig4} shows the excellent agreement between their numerical
simulation and our similarity calculation.

\subsection{ Analytic Approximation}

Following FTB we compare the results of the self-similar 
champagne flows with a simple analytic approximation. 
Assume that the pressure driving the  shock is uniform,
\be
P(t) = a^2 \overline{\rho}(t),
\ee
where $\overline{\rho}(t)$ is the mean density at a given time, i.e.
\be
\overline{\rho}(t)=
{\int_{0}^{r_s} 4 \pi r^2 \rho(r,t) dr
\over \int\limits_{0}^{r_s} 4 \pi r^2 dr}
= \left({3 \over 3 - n}\right) \rho[r_s(t)] 
\quad \mbox{with}\quad n<3,
\label{rmean}
\ee
where $r_s(t)$ is the instantaneous position of the shock front.
This must be equal to the (strong shock) post-shock pressure given by
\be
 P(t) =(u_s-u_u)^2 \rho[r_s(t)], 
\ee
where $u_u$ is the upstream velocity at the position of the shock front.
Equating these two pressures, one obtains 
\be
x_s = { u_s \over a}= \sqrt{{3 \over
{3-n}}} +  {u_u \over a},
\label{vs}
\ee
which is equivalent to eq. (24) of FTB when $u_u = a$.  Eq. (\ref{vs})
diverges as $n \to 3$ because, in this limit, the mass inside any
radius diverges if the origin is included (eq.~\ref{rmean}).

Figure \ref{fig4} shows $\log (x_s) $ vs. $n$ for the self-similar
models. The long dashed line corresponds to eq.~(\ref{vs}) where we
have used $u_u = a$ (as FTB).  The dotted line instead is
eq.~(\ref{vs}) with $u_u = 0$.  The difference is due, in part, to the
pressure driving the shock being not exactly uniform as was assumed in
the approximation above.  The deviation from uniform density was
commented upon in \S\S 3 and 5 when we discussed the approximate nature
of the fits provided by the dashed lines in the lower panels of Figures
\ref{fig2} and \ref{fig3}.
 
To avoid the divergence at the origin, one could integrate from the
radius of influence $r_0$ (or include a uniform core as considered by
FTB).  In that case, the density can be evaluated for all $n$ and is
given by
\be
\overline{\rho}(t)
= f[r_s(t)] \rho[r_s(t)], 
\ee
where
\be
f[r_s(t)]
= \left\{ 
\begin{array}{ll}
{3 \over 3 - n }
{1 \over \Omega}
\left (1 - \left [{ r_0 / r_s(t) }\right ]^{3-n} \right)
& \mbox{if $n \neq 3$} \\
{1 \over \Omega} \ln\left[{r_s(t) / r_0}\right]^3 & \mbox{if $n=3$} 
\end{array}
\right.
\ee
and $\Omega = 1 - [r_0/r_s(t)]^3$ is a volume factor.
The shock velocity is then
\be
u_s=
a \sqrt{f[r_s(t)] \, }
+ u_u.
\label{vsfinite}
\ee
For $n <3$, $f(r_s) \rightarrow 3/(3-n)$ when $r_s \gg r_0$; thus, the shock
reaches the asymptotic
constant value given by eq.~(\ref{vs}). For the case $ n \geq 3$, the shock
accelerates to $\infty$ as $r_s \to \infty$.  The latter divergence
arises only because a shockwave of finite outward momentum is allowed
to run into spherical shells of ever decreasing mass.

Finally, as discussed in Appendix A and B, even though the expansion
velocities become very large as $n \rightarrow 3$, adiabatic cooling
never offsets photoionization heating and the gas remains isothermal.
We also show that for the scales of interest in molecular clouds, even
if the shock velocity becomes large, the assumption of an isothermal
shock remains valid.

\section{Discussion}

Compact \hii regions with velocity gradients indicative of champagne
flows are characterized by values of the emission measure (EM) in the
range $10^6$~cm$^{-6}$~pc $ \simlt \mbox{EM} \simlt 10^8$~cm$^{-6}$~pc
(e.g. Garay et al. 1994).  The results of FTB and the self-similar
models presented here show that when a molecular core is ionized and
heated out of equilibrium, steep density gradients characteristic of
star-forming regions produce shocks that travel at constant velocity and
accelerate the ionized gas to supersonic speeds.  This phase of
supersonic expansion of the ionized gas poses a lifetime problem for
champagne flows, more severe than the one pointed out by Wood \&
Churchwell (1989) in the case of ultracompact \hii regions.

The emission measure of the self-similar models presented here at time
$t=0^+$, just after ionization of the cloud core is 
\be
{\rm EM}_0 \equiv\int_{r_0}^\infty n_e n_p dr
=\left( {1 \over 2n-1} \right) \left({K \over 2 \mu_i m_H}\right)^2 
r_0^{1-2n}.
\label{em0}
\ee
As we discussed in \S~\ref{intro}, if $K < K_{\rm cr}$ the \hii region
is density bounded and therefore able to develop a champagne flow.
With $K_{\rm cr}$ defined by eq.~(\ref{Kcrit}), and our fiducial values
of the physical parameters for a 25 $M_\odot$ star,
this condition results in the following
upper limit on the EM:
\be
{\rm EM}_0  < \left({2n-3 \over 2n-1}\right)
{\dot N_\ast \over 4\pi\alpha_2 r_0^2} \sim 4 \times 10^7~\mbox{cm$^{-6}$~pc},
\ee
in the middle of the range indicated by the observations.  However,
inserting the expression of $\rho(r,t)$ given by eq.~(\ref{plaw}) into
the eq.~(\ref{em0}), one easily obtains $\mbox{EM}(t)\propto t^{1-2n}$
for $t\gg r_0/a$. Thus, the expansion of the champagne flows results in
a rapid decrease of the emission measure, and in a short time the 
source fades away from observational classification as a compact \hii
region.  Inclusion of the effects of a fast stellar wind would only
exacerbate the situation.

These estimates show that a continuous source of ionized mass is
required to keep up the density of the expanding champagne flow.
Photoevaporation of circumstellar disks (Hollenbach et al. 1994;
Richling \& Yorke 1997) or
mass loading by the photoevaporation and/or hydrodynamic ablation
of remnant neutral globules
surrounding the central star (e.g. Lizano et al. 1996; Redman et al. 1996)
are natural solutions
to maintain the high observed emission measures in champagne flows and,
in general, in ultracompact \hii regions. In particular, proplyds in
Orion, sites of low mass star formation, are known to have appreciable
mass loss rates that can mass load the stellar wind (St\"orzer \&
Hollenbach 2000; Garc\'\i a-Arredondo et al. 2002).  Revealed O and B
stars show a lower frequency of surrounding disks than stars of
spectral type A and later (Natta et al.~2000; Lada et al.~2000),
suggestive of a picture where the disks are photoevaporated during an
earlier phase of evolution, perhaps as ultracompact \hii regions.  As a
first attempt to model quasi-spherical ultracompact \hii regions on
this basis, it would be interesting to extend the methods of this paper
to include a continuous source of ionized gas (due to photoevaporation
of a disk or neutral globules into a mass-loaded stellar wind) that emanates at a
prescribed rate and speed from the origin.

\section{Conclusions}

We have obtained self-similar champagne flow solutions for the
expansion of power law  gas density distributions after the gas has
been uniformly heated out of mechanical equilibrium by the birth of a
star at the center of a molecular cloud core. These solutions attach
via a shock to upstream breeze solutions.

In the case of the isothermal sphere with $\rho \propto r^{-2}$, the
``inside-out expansion''  found by Tsai \& Hsu (1995) is the limit of
the family of self-similar outflow solutions when the sound speed $a_2$
of the \hii region is the same as the sound speed $a_1$ of the original
molecular cloud core.  The case $(a_1/a_2)^2=1$ must include the effect
of the self-gravity of the gas, and the outflow solution then attaches
to the unperturbed static SIS upstream of the shock.  Another member of
this family, for $(a_1/a_2)^2=0.75$, is a time reversed piece of the
L-P collapse solution, with a shock allowing the upstream solution to
have a correct outflow (breeze) asymptotic behavior.

For the high values of the gas temperature expected after the passage
of an ionization front, $(a_1/a_2)^2 \ll 1$, and the self-gravity of the
gas can be neglected.  The solution then approaches a shape invariant
form.  In the approximation that the self-gravity of the ionized gas
can be ignored, we computed the self-similar champagne flows of \hii
regions formed in molecular clouds characterized by power law density
distributions with exponents $3/2 < n < 3$.  These self-similar
solutions behave as in the numerical models of FTB: in the ``champagne
phase'' a shock moves with a constant speed into the ionized medium
with a shock speed that increases with increasing density gradient.
The speed of the shock diverges as $n \rightarrow 3$ because the mass
of the driving gaseous piston diverges, if the origin is included.
Instead, if the origin is excluded, the shock front velocity reaches an
asymptotic constant value for $n<3$. For $n \ge 3$ the shock
accelerates to infinite velocity as $r_s \rightarrow \infty$, but only
because finite outward momentum is inputted into spherical shells of
ever decreasing mass.  These results may help explain astrophysical
champagne flows where expansion velocities are seen that are
considerably larger than the sound speed $a_2\simeq 10$ km s$^{-1}$
associated with conventional \hii regions.  (Driving by fast stellar
winds may contribute to the perceived motions.) Despite the large
expansion velocities produced in the case of the steepest pressure
gradients ($n \rightarrow 3$), we show in the appendices that the
isothermal assumption for the gas and for the shock are valid for the
scales of interest in molecular clouds.

The supersonic expansion of the ionized gas creates a severe lifetime
problem for champagne flows.  A natural solution is the
photoevaporation of circumstellar disks and/or remnant neutral 
globules which would help maintain the
high observed emission measures in these sources.

\acknowledgements

We thank Luis Rodr\'\i guez, Malcolm Walmsley and Frank Wilkin 
for useful comments on the manuscript.  
F. S.  acknowledges support in the United States from the
NSF and NASA, and in Taiwan from the National Science Council.  S. L.
and J. C. acknowledge support from DGAPA/UNAM and CONACyT.  
D. G. acknowledges support from grant COFIN-2000 and warm 
hospitality from the Instituto de Astronom\'\i a, UNAM in Morelia.

\appendix
\section{Adiabatic Cooling vs. Photoionization Heating}
\label{apA}

For an expanding nebula, the equilibrium temperature is given by the
balance between photoionization heating $\Gamma_{\rm ph}$ and radiative
cooling $\Lambda_{\rm rad}$ and adiabatic cooling $\Lambda_{\rm ad}$.

We can estimate when adiabatic cooling becomes important by setting
\be
\Gamma_{\rm ph} = \Lambda_{\rm ad}.
\ee
Assuming ionization equilibrium and the ``on the spot'' approximation,
$\Gamma_{\rm ph}$ is given by
\be
\Gamma_{\rm ph} \simeq n_e n_p \alpha_2 {3 \over 2} k T_\ast, 
\ee
where $T_\ast$ is the stellar temperature (e.g. Osterbrock 1989).
The rate of adiabatic cooling (for a constant mass) is given by
\be
\Lambda_{\rm ad} = {P \over V} {d V \over d t} = a^2
\left \vert { d \, \overline{\rho}
\over d t } \right\vert .   
\ee
Then, the condition for adiabatic cooling to win over
photoionization heating can be written as
\be
{ t_{\rm rec} \over t_\rho } \geq {3 \over 4}  {T_\ast \over
T},
\ee
where $t_{\rm rec} \equiv 1 / n_e \alpha_2 $ is the recombination
timescale, and $t_\rho \equiv \overline{\rho} / \vert {d
\overline{\rho} /d t }\vert $ is the timescale for the decrease in the
density.

Taking $T_\ast / T \simeq 4$, the critical ratio of timescales
is 
\be
{t_{\rm rec} \over t_\rho} \simeq 3.  
\label{critra}
\ee

On the other hand, using eq.~(\ref{plaw}) and (\ref{rmean}) this ratio
can be written as
\be
{ t_{\rm rec} \over t_\rho}
= 2 n \left({3 - n \over 3 }\right) 
{ \mu_i m_H a \over K \alpha_2} x_s r_s^{n-1}.
\ee

Using the critical value for the ratio of timescales (\ref{critra}),
one can solve this equation for a critical radius $r_{\rm cr}$
beyond which the isothermal assumption breaks down. For convenience,
one can write the density constant in eq. (\ref{powerlaw}) as $K = \mu
\, m_H n_0 \, r_0^n$, where $\mu$ is the mean molecular weight and
$n_0$ is the number density at the distance $r_0$. Then, the critical
radius can be written as
\be
r_{\rm cr} = r_0
\left [{\alpha_2  \mu n_0 r_0 \over 2 \mu_i a x_s}
\left ( {3 \over 3-n} \right ) \left ( {3 \over n} \right )
\right ]^{1 / (n-1)},
\ee
Finally, for $\alpha_2\simeq 2.6 \times 10^{-13}$~cm$^3$~s$^{-1}$ and
the fiducial parameters $n_0 \simeq 10^4$ cm$^{-3}$ $r_0\simeq 10^4$ AU, $a \simeq 10 \,
{\rm km \, s^{-1}}$,  and $\mu \simeq 2 $, typical of massive molecular
cloud cores, the critical radius becomes
\be
r_{\rm cr} \simeq r_0 \left[ 780
\left ( {3 \over 3-n} \right )^{1/2} \left( {3 \over n}\right) \right ]^{1 /(n-1)},
\ee
where we used eq.~(\ref{vs}) with $u_u = 0 $ to obtain the last
expression.

This critical radius is large compared to $r_0$ and the sizes of
molecular cloud cores where the stars are formed. Thus, adiabatic
cooling never dominates over photoionization heating and the assumption
of an isothermal gas is justified.

\section{Isothermal Shock}

We now examine the assumption that the shock is isothermal in the case
when the shock speed is large ($ n \rightarrow 3$). The post shock
temperature is
\be
T_d \simeq {3 \over 16} {\mu_i \, m_H \over k} u_s^2 = 1.2 \times 10^3 
\left ({u_s \over 10 \, {\rm km \, s^{-1}} } \right)^2 \, {\rm  K}. 
\label{ts}
\ee
For large shock velocities the gas behind the shock will not be able to
cool efficiently, thus the cooling time may become larger than the
expansion time, $t_{\rm cool} > t_{\rm exp} = r_s / u_s$, and the shock
may become energy conserving.

The expansion time is
\be
t_{\rm exp} =  1.5 \times 10^{11} \left({r \over 10^4 \, {\rm AU} } \right) 
\left({ u_s \over 10 \, {\rm km \, s^{-1}} } \right)^{-1}~\mbox{s}.
\ee

The cooling time is given by
\be
t_{\rm cool} =  {3 \over 2} { k T_d \over n_d \Lambda_{\rm rad} },
\ee
where $T_d$ is the gas post-shock temperature, $n_d$ is the gas
post-shock number density and $\Lambda_{\rm rad}$ is the cooling function 
(in erg~cm$^3$~s$^{-1}$).
Substituting the post-shock temperature
(\ref{ts}), using eq.~(\ref{jump}) for the post-shock density and
assuming that the upstream flow is as rest, the cooling time can be
rewritten as
\begin{eqnarray}
t_{\rm cool}
& = & {9 \, \mu_i^2 \, m_H \, a^2 \over 32 \, \mu} {1 \over n_0 \, \Lambda_{\rm rad} }
\left( {r_s \over r_0} \right)^n \nonumber \\
& \simeq & 1.4 \times 10^4
\left({\Lambda_{\rm rad} \over 10^{-23}~\mbox{erg~cm$^3$~s$^{-1}$}} \right)^{-1}
\left( {r_s \over 10^4 \, {\rm AU}} \right)^n {\rm s},
\end{eqnarray}
where we have again used the expression for $K$ defined in Appendix~A.

Thus, the ratio is
\be
{t_{\rm cool} \over t_{\rm exp}} \simeq 10^{-7}
\left({ \Lambda_{\rm rad} \over 10^{-23}~\mbox{erg~cm$^3$~s$^{-1}$}}
\right)^{-1}
\left( {r_s \over 10^4 \, {\rm AU}} \right)^{n-1} 
\left({ u_s \over 10 \, {\rm km \, s^{-1}} } \right).
\ee

This equation is expressed in terms of typical values of the cooling
function (e.g. Dalgarno \& McCray 1972) and shows that the ratio of
cooling time to expansion time is small.  Thus, the assumption of an
isothermal shock will remain valid for the scales relevant in molecular
clouds.

\clearpage

\begin{figure}
\plotone{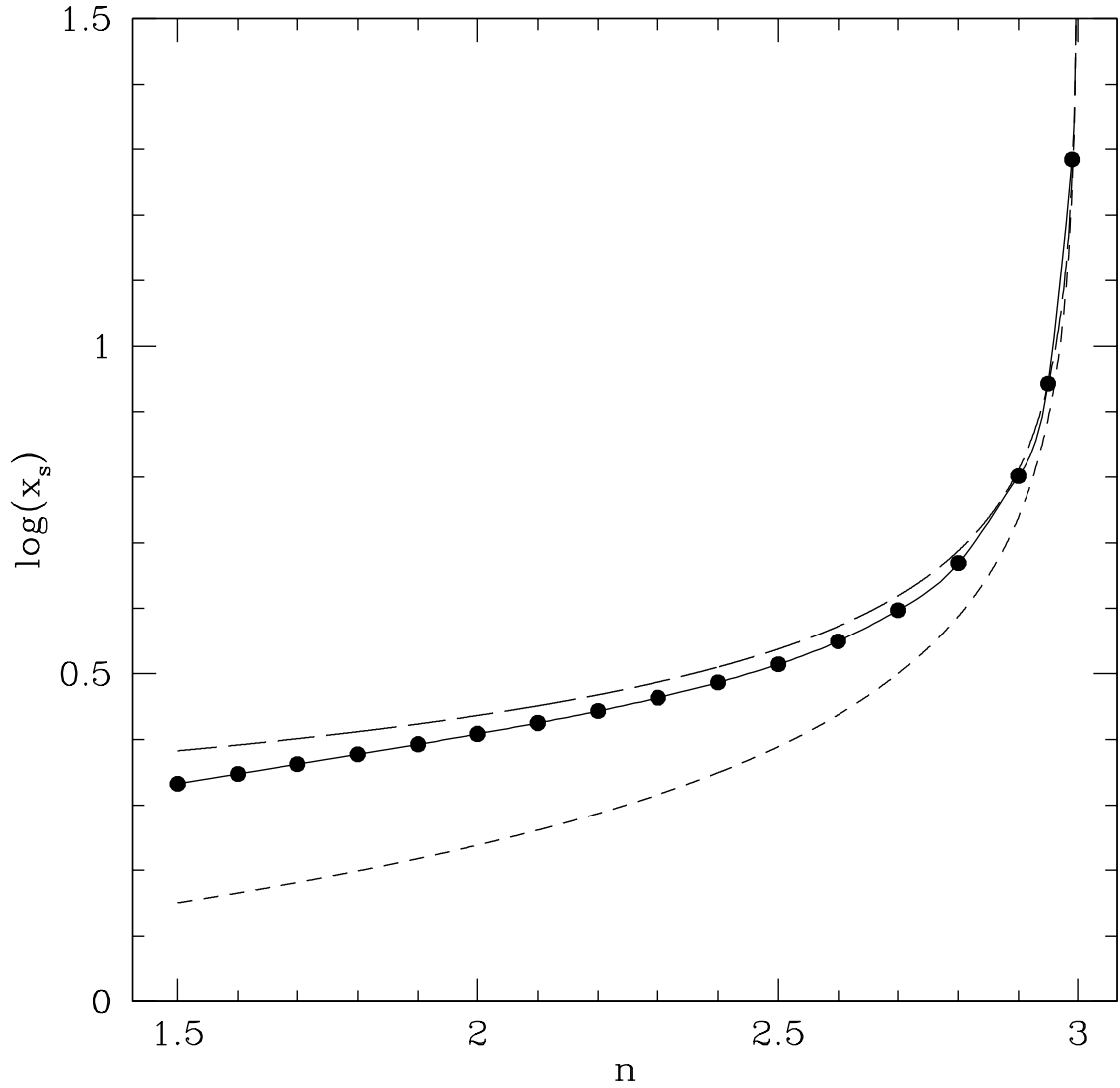}
\caption{ Plot of $x_s$ vs. $n$ showing the divergence of the shock
position as $n \rightarrow \infty$. The {\it dots} 
represent the exact self-similar solutions. The {\it long-dashed
curve} and the {\it short-dashed curve} correspond to the analytic
approximation eq.~(\ref{vs}) with $u_u = a$ and $u_u = 0$,
respectively.}
\label{fig5}
\end{figure}


\begin{thebibliography}{}

\bibitem[Bodenheimer et al. (1979)]{B79} Bodenheimer, P., 
Tenorio-Tagle, G., \& Yorke, P. 1979, ApJ, 233, 85

\bibitem[Caselli \& Myers]{CM95}Caselli, P., \& Myers, P. 1995, \apj, 446, 665

\bibitem[Dalgarno \& McCray 1972]{DM72} Dalgarno, A., \& McCray, R. A. 1972, ARA\&A, 10, 375

\bibitem[Franco et al. 2000]{FKHT2000} Franco, J., Kurtz, S., 
Hofner, P., \& Testi, L. 2000, ApJ, L143

\bibitem[FTB]{f90} Franco, J., Tenorio-Tagle, G., \& Bodenheimer, P. 1990,
 ApJ, 349, 126 (FTB)

\bibitem[Galli et al. (1999)]{Ga99} Galli, D., Lizano, S., Li., Z.-Y.,
Adams, F. C., \& Shu, F. H. 1999,
 ApJ, 521, 630

\bibitem[Garay \& Lizano 1999]{GL99} Garay, G., \& Lizano, S. 1999, PASP, 111, 1049

\bibitem[Garay et al. 1994]{G94} Garay, G., Lizano, S., \& G\'omez, Y. 1994,
ApJ, 429, 268

\bibitem[Garay \& Rodr\'\i guez 1990]{GR90} Garay, G., \& 
Rodr\'\i guez, L.~F. 1990, ApJ, 362, 191

\bibitem[Garc\'\i a-Arredondo et al. 2002]{GAH02} Garc\'\i a-Arredondo, F.,
Arthur, J., \& Henney, W. 2002, Rev. Mex. Astron. Astrof., 38, 51

\bibitem[Hatchell et al. (2000)]{Ha2000} Hatchell, J., Fuller, G. A., Millar, T. J.,
Thompson. M. A., \& Macdonald, G. H. 2000, A\&A, 357, 637

\bibitem[Hennebelle (2001)]{H01}
Hennebelle, P. 2001, A\&A, 378, 214

\bibitem[Hunter (1977)]{H77} Hunter, C. 1977, ApJ, 218, 834

\bibitem[Hollenbach et al. 1994]{hal} Hollenbach, D., Johnstone, D., Lizano, S.,
\& Shu, F. H. 1994, ApJ, 428, 654

\bibitem[Keto et al. 1995]{K95} Keto, E. R., Welch, W. J., Reid, M. J.,
\& Ho, P. T. P. 1995, ApJ, 444, 765

\bibitem[Lada et al. (2000)]{lal} Lada, C. J., Muench, A. A., Haisch, K. E., 
Lada, E., Alves, J. F., Tollestrup, E. V., \& Willner, S. P. 2000, 
AJ, 120, 3162

\bibitem[Larson 1969]{L69} Larson, R. B. 1969, MNRAS, 145, 271

\bibitem[Lebron et al. (2001)]{Le2001} Lebr\'on, M., Rodr\'\i guez, L. F., 
\& Lizano, S.  2001, ApJ, 560, 806

\bibitem[Lizano et al. 1996]{LCGH96} Lizano, S., Cant\'o, J., Garay, G.,
\& Hollenbach, D. 1996, ApJ, 739

\bibitem[Lumsden \& Hoare]{LH1996} Lumsden, S. L., \& Hoare, M. G. 1996,
ApJ, 464, 272

\bibitem[McKee \& Tan 2002]{mct} McKee, C., \& Tan, J. C. 2002, Nat. 416, 59

\bibitem[Natta et al. 2000]{ngm} Natta, A., Grinin, V. P., \& Mannings, V.
2000, in {\it Protostars and Plantes IV}, eds. V. Mannings, A. P. Boss, \&
Russell, S. S. (Tucson: The university of Arizona Press), p. 559

\bibitem[Newman \& Axford (1968)]{NA68}Newman, R. C., \& Axford, W. I. 
1968, ApJ, 153, 595

\bibitem[Osorio et al. 1999]{old} Osorio, M., Lizano, S., \& D'Alessio, P. 1999,
ApJ, 525 808

\bibitem[Osterbrok (1989)]{O89} Osterbrok, D. E. 1989, {\it Astrophysics of
Gaseous Nebulae and Active Galactic Nuclei} (Mill Valley:  University Science Books)

\bibitem[Penston 1969]{P69} Penston, M. V. 1969, MNRAS, 144, 425

\bibitem[Redman et al. (1996)]{RWD96} Redman, M. P., Williams, R. J. R., \&
Dyson, J. E. 1996, MNRAS, 280, 661

\bibitem[Richling \& Yorke (1997)]{RY97} Richling, S., \& Yorke, H.W. 1997, A\&A, 327, 317

\bibitem[Shu (1977)]{S77} Shu, F. 1977, ApJ, 214, 488

\bibitem[St\"orzer \& Hollenbach 2000]{SH02} St\"orzer, H. \& Hollenbach,
D. 1999, ApJ, 539, 751

\bibitem[Str\"omgren 1939]{Str39} Str\"omgren, B. 1939, ApJ, 89, 526

\bibitem[Tsai \& Hsu (1995)]{TH95} Tsai, J. C., \& Hsu, J. J. L. 1995,
ApJ, 448, 787

\bibitem[Vacca et al. (1996)]{vgs} Vacca, W. D., Garmany, C. D., \& Shull, J. M.
1996, ApJ, 460, 914

\bibitem[van der Tak et al. 2000]{v02} van der Tak, F. F. S., van Dishoeck, E. F.,
Evans, N. J. II, \& Blake, G. A. 2000, ApJ 537, 283

\bibitem[Whitwort \& Summers (1985)]{ws85}
Whitworth, A., \& Summers, D. 1985, MNRAS, 214, 1

\bibitem[Wood \& Churchwell 1989]{wc89} Wood, D. O. S., \& Churchwell, E. 1989,
ApJSS, 69, 831

\bibitem[Yorke et al (1983)]{Y83} Yorke, H. W., Tenorio-Tagle, G., \&
Bodenheimer, P. 1983, A\&A, 127, 313

\end{thebibliography}
\end{document}